\documentclass[runningheads,a4paper]{llncs}

\usepackage{amsfonts,amssymb,amsmath}
\usepackage{natbib}

\usepackage{wrapfig}
\usepackage{llncsdoc}
\usepackage{epsfig,psfrag}
\usepackage{latexsym}
\usepackage{longtable}
\usepackage{tikz,pgf}
\usepackage{subfig}
\usepackage{wrapfig}

\usepackage{graphicx}
\usepackage{epstopdf}
\usepackage{color}

\newcommand{\bqn}{\begin{eqnarray}}
\newcommand{\eqn}{\end{eqnarray}}
\newcommand{\bq}{\begin{eqnarray*}}
\newcommand{\eq}{\end{eqnarray*}}
\usepackage[ruled,vlined]{algorithm2e}

\usepackage{url}
\usepackage{hyperref}
\hypersetup{colorlinks,%
citecolor=black,%
filecolor=blue,%
linkcolor=red,%
urlcolor=blue,%
pdftex}

\newcommand{\blue}[1]{{\color{blue} #1}}

\begin{document}

\title{Gaussian Kernel Smoothing}
 \titlerunning{Gaussian Kernel Smoothing}
 
\author{Moo K. Chung
 }
\institute{
University of Wisconsin-Madison, USA\\
\vspace{0.3cm}
\blue{\tt mkchung@wisc.edu}
}
\authorrunning{Chung}

\maketitle

\begin{center}
July 30, 2012
\end{center}

\pagenumbering{arabic}

\index{smoothing ! Gaussian kernel}
\index{Gaussian kernel smoothing}

Image acquisition and segmentation are likely to introduce noise. Further image processing such as image registration and parameterization can introduce additional noise. It is thus imperative to reduce noise measurements and boost signal. In order to increase the signal-to-noise ratio (SNR) and smoothness of data required for the subsequent random field theory based statistical inference, some type of smoothing is necessary \citep{kiebel.1996}. Among many image smoothing methods, {\em Gaussian kernel smoothing} has emerged as a de facto smoothing technique among brain imaging researchers  due to its simplicity in numerical implementation \citep{kovacic.1999, perona.1990}. Gaussian kernel smoothing also increases statistical sensitivity and statistical power as well as Gausianness. Gaussian kernel smoothing can be viewed as weighted averaging of voxel values. Then from  the central limit theorem, the weighted average should be more Gaussian. This paper reproduces Chapter 4 of book \citet{chung.2013.SCM} with error corrections and additional materials. The MATLAB codes and sample data used in the paper can be downloaded from \url{http://brainimaging.waisman.wisc.edu/~chung/BIA}.


\section{Kernel Smoothing}
\index{kernel ! Gaussian}
\index{kernel smoothing ! Gaussian}

\begin{figure}[t]
\includegraphics[width=0.8\linewidth]{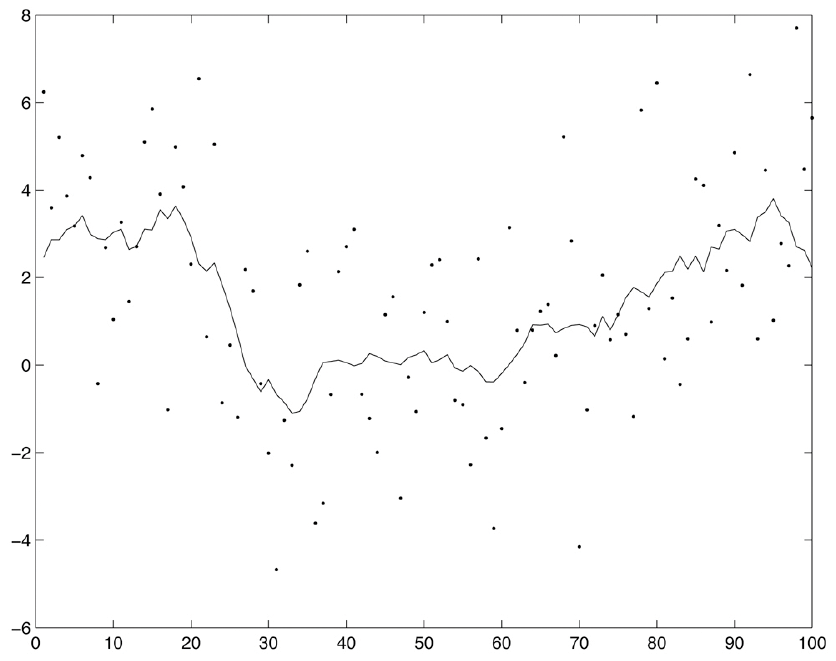}
\centering
\caption{Simulated noise $N(0,2^2)$ is added to signal  $\mu(t) = (t-50)^2/500$. Gaussian kernel smoothing is applied with bandwidth 10.}
\label{fig:kernel1D}
\end{figure}

Kernel smoothing is the most widely used image smoothing technique in brain image analysis. Consider the integral transform
\bq 
Y(t) = \int K(t,s)X(s) \; ds, \label{eq:smoothing-linearsystem}
\eq
where $K$ is the {\em kernel} of the integral. Given the input signal $X$, $Y$ represents the output signal. The smoothness of the output depends on the smoothness of the kernel.  We assume the kernel to be unimodal and isotropic. When the kernel is isotropic, it has radial symmetry and should be invariant under rotation. So it has the form 
$$K(t,s) = f(\|t-s\|)$$
for some smooth function $f$. Since the kernel only depends on the difference of the arguments, with the abuse of notation, we can simply write $K$ as
$$K(t,s)= K(t-s).$$
We may further assume the kernel is normalized such that
$$\int K(t) \; dt =1.$$ 
With this specific form of kernel $K$, (\ref{eq:smoothing-linearsystem}) can be written as
\bq 
Y(t) =  K*X(t) = \int K(t-s)X(s) \; ds \label{eq:smoothing-KS}. 
\eq
The integral transform (\ref{eq:smoothing-KS}) is called {\em kernel smoothing}. We may assume further that $K$  is dependent on {\em bandwidth} $\sigma$. The bandwidth will determine the spread of kernel weights such that  
\bqn \lim_{\sigma \to \infty} K(t,s;\sigma) &=& 1 \\
 \lim_{\sigma \to 0} K(t,s;\sigma) & =&  \delta(t-s), \eqn
where $\delta$ is the Dirac-delta function \citep{berline.1991,dirac.1981}. The Dirac-delta function is a special case of {\em generalized functions} \citep{gelfand.1964,stakgold.2000}. The Dirac-delta function is usually traditionally defined as
$$\delta(t)=0  \mbox{ if } t \neq 0, \quad \delta(t) = \infty  \mbox{ if } t = 0.$$
$$\int_{-\infty}^{\infty} \delta(t) \;dt =1. $$
The Dirac delta function is also referred to as the {\em impulse function} in literature. Figure \ref{fig:kernel1D} illustrates 1D Gaussian kernel smoothing in simulated data. One of the important properties of Dirac delta function is 
$$\int_{-\infty}^{\infty} f(t) \delta(t) \; dt = f(0).$$
Then it follows the one of most important operational characteristic of Dirac delta:
\bqn K* \delta(t) = \int_{-\infty}^{\infty} K(t-s) \delta(s) \; ds = K(t). \eqn

\section{Gaussian Kernel Smoothing} 
\index{kernel ! isotropic Gaussian}
\index{kernel smoothing ! Gaussian}

All brain images are inherently noisy due to errors associated with image acquisition. Compounding the image acquisition errors, there are errors caused by image registration and segmentation. So it is necessary to smooth out the segmented images before any statistical analysis is performed to boost statistical power. Among many possible kernel smoothing methods \citep{kovacic.1999, perona.1990}, {\em Gaussian kernel smoothing} has emerged as a de facto smoothing technique in brain imaging. 

\index{Gaussian kernel}
\index{kernel ! Gaussian}

The {\em Gaussian kernel} in 1D is defined as
$$K(t) = \frac{1}{\sqrt{2\pi}}e^{t^2/2}.$$
Let's scale the Gaussian kernel $K$ by the bandwidth $\sigma$:
$$K_{\sigma}(t) = \frac{1}{\sigma}K\Big(\frac{t}{\sigma}\Big).$$
This is the density function of the normal distribution with mean $0$ and variance $\sigma^2$.

The $n$-dimensional isotropic Gaussian kernel  is defined as the product of $n$ 1D kernels. Let $t=(t_1,\cdots,t_n)^{\top} \in \mathbb{R}^n.$ Then the $n$-dimensional kernel is given by
\bq K_{\sigma}(t) &=& K_{\sigma}(t_1)K_{\sigma}(t_2) \cdots K_{\sigma}(t_n)\\
                    &=&    \frac{1}{(2\pi)^{n/2}\sigma^n} \exp \Big( 
\frac{1}{2\sigma^2} \sum_{i=1}^n t_i^2 \Big).
\eq
Subsequently, $n$-dimensional isotropic Gaussian kernel smoothing $K_{\sigma}*X$ can be done by applying 1-dimensional smoothing $n$ times in each direction by factorizing the kernel as
\bq
K_{\sigma}*X(t) &=& \int K_{\sigma}(t -  s)X( s)\; ds\\
&=& \int K_{\sigma}(t_1-s_1)K_{\sigma}(t_2-s_2) \cdots K_{\sigma}(t_n-s_n) X(s) \;d s\\
&=& \int       K_{\sigma}(t_1-s_1) \cdots K_{\sigma}(t_{n-1}-s_{n-1}) \; ds_1\cdots ds_{n-1}   \\
&&\times \int K_{\sigma}(t_n-s_n)X(s)\;ds_n.
\eq
Note that $K_{\sigma}*X$ is the scale-space representation of image $X$ first introduced in \citep{witkin.1983}. Each $K_{\sigma}*X$ for different values of $\sigma$ produces a blurred copy of its original. The resulting scale-space representation from coarse to fine resolution can be used in multiscale approaches such as hierarchical searches and image segmentation \citep{lindeberg.1994,poline.1994.TMI,poline.1995,worsley.1996,worsley.1996.hbm}.

\section{Effective FWHM}
\index{kernel ! FWHM}
\index{full width at half maximum (FWHM)}

The {\em bandwidth} $\sigma$ defines the spread of kernel. In brain imaging, the spread of the kernel is usually measured in terms of the {\em full width at the half maximum} (FWHM) of Gaussian kernel $K_{\sigma}$, which is given by $2\sqrt{2 \ln 2} \sigma$. This can be easily seen by representing the kernel using the radius $r^2 = x_1^2 + \cdots x_n^2$:
$$K_{\sigma} (r) = \frac{1}{(2\pi)^{n/2}\sigma^n} \exp \Big(-\frac{r^2}{2\sigma^2}\Big).$$   
The peak of the kernel is $K_{\sigma}(0)$. Then the {\em full width at half maximum} (FWHM) of the peak  is given by 
$$\mbox{ FWHM} = 2 \sqrt{2 \ln 2} \sigma.$$
The FWHM increases linearly as $\sigma$ increases in Euclidean space. 

\index{kernel ! full width at half maximum}
\index{full width at half maximum (FWHM)}
\index{effective-FWHM}

Once we smooth the image with kernel $K_{\sigma}$, the smoothness of the image changes. The smoothness of image after smoothing can be measured in terms of {\em effective FWHM}. The unbiased estimator of eFWHM is first introduced in \citep{worsley.1999}, where it is estimated along edges in the lattice. Label the two voxels at the end of an edge by 1 and 2. Let the length of edge be $\Delta x$. Suppose there are $n$ images in a group. Let $r_{ij}$ denote the residual for the $i$-th image at voxel $j$. The normalized residuals at the two ends are
$$u_{ij}  = \frac{r_{ij}}{\sqrt{ \sum_{i=1}^n r_{ij}^2}}.$$
The roughness of the noise is defined as the standard deviation of the derivative of the noise divided by the standard deviation of the noise itself. 
Let
$$\Delta u =  \sqrt {\sum_{i=1}^{n} (u_{i1} - u_{i2})^2}.$$ 
Then an unbiased estimator of the roughness is given by
$$\lambda = \frac{\Delta u}{\Delta x}.$$
Then the eFWHM along the edge is given by 
$$e\mbox{FWHM} = \frac{\sqrt {4 ln 2}} \lambda.$$

The effective-FWHM is often used in the random field theory based on multiple comparisons correction through the {\tt fMRISTAT} package \citep{worsley.2004}.

\section{Numerical Implementation}
\label{sec:smooth-MATLAB}

We present 1D and 2D Gaussian kernel smoothing here as illustrations using MATLAB. The codes and relevant image examples can be downloaded from
\url{http://brainimaging.waisman.wisc.edu/~chung/BIA/download/matlab.v1}. The corresponding Matlab script is 
{\tt chapter04-smoothing.m}.

\subsection{Smoothing Scalar Functions}

Consider a noise 1D functional signal
$$Y(t) = (t-50)^2/500 + \epsilon(t),$$
where $\epsilon(t)$ is distributed as $N(0, 2^2)$ at each point $t$. At each integer point between 0 and 100, we have noisy measurement (Figure \ref{fig:kernel1D}). We are interested in smoothing out the noisy measurement and estimating the underlying smooth signal. Using {\tt inline} function, we define kernel {\tt K} and kernel smoothing is performed by convolution.

\begin{verbatim}
K=inline('exp(-(x.^2)/2/sigma^2)');
dx= -5:5;

sum(K(10,dx))
weight=K(10,dx)/sum(K(10,dx))
sum(weight)

t=1:100;
mu=(t-50).^2/500;
noise= normrnd(0, 2, 1,100);
Y=mu + noise
figure; plot(Y, '.');

smooth=conv(Y,weight,'same');
hold on; plot(smooth, 'r'); 
\end{verbatim}

\subsection{Smoothing Image Slices}
\begin{figure}[t]
\includegraphics[width=1\linewidth]{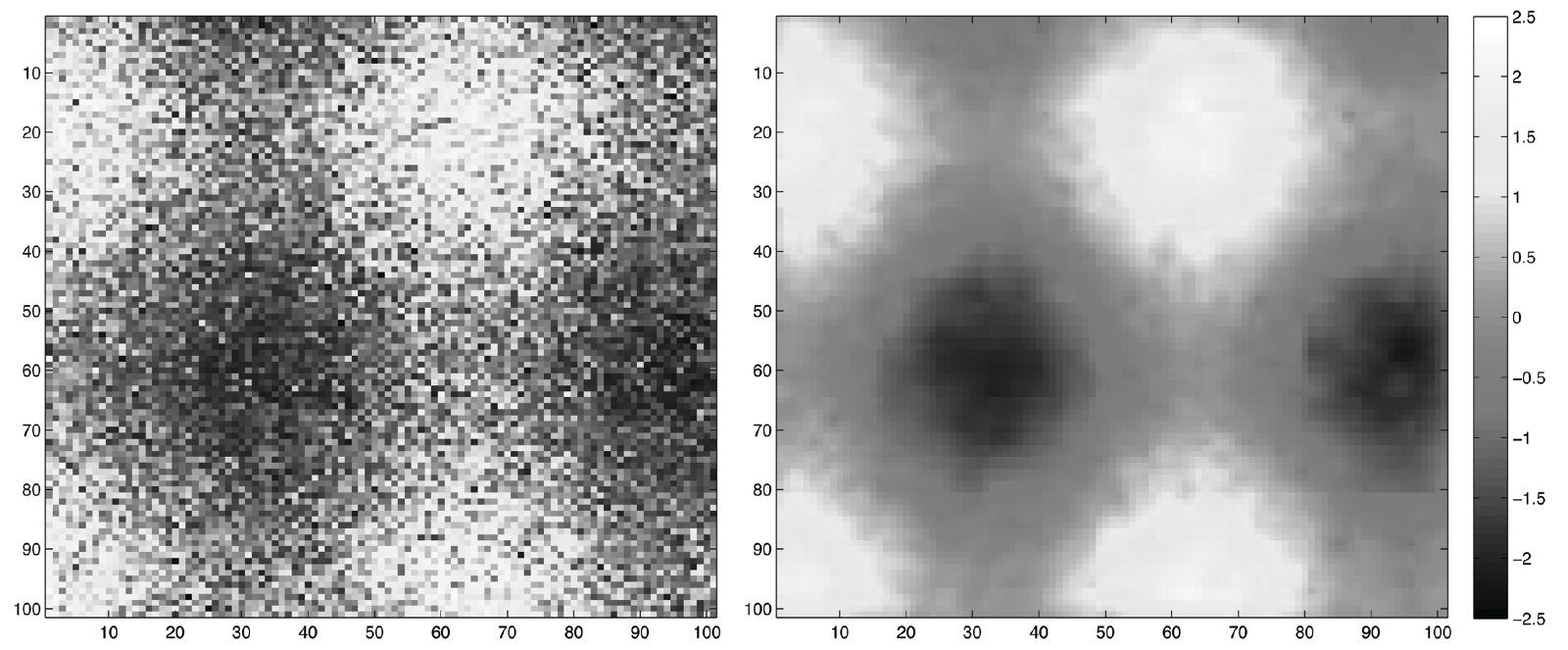}
\centering
\caption{Left: simulated noise $N(0,0.4^2)$ is added to signal  $\mu(t_1,t_2) = \cos (10 t_1) +\sin(10 t_2)$. Right: Gaussian kernel smoothing is applied with bandwidth 1.}
\label{fig:kernelsmoothing}
\end{figure}

Suppose that noisy observation is obtained at each grid
point $(\frac{i}{100},\frac{j}{100}) \in [0,1]^2.$ This basically forms a 2D image. 
The signal $\mu$ is assumed to be
$$ \mu(t_1,t_2) = \cos (10 t_1) +\sin(10 t_2).$$
The signal is contaminated by Gaussian white noise
$\epsilon \sim N(0,0.4^2)$.
Using {\tt meshgrid} command, we generate $101 \times 101$ 2D grid points. The grid coordinates are stored in {\tt px} and {\tt py}, which are $101 \times 101$ matrix. Figure \ref{fig:kernelsmoothing} shows the simulated image $Y=\mu + \epsilon$. 

\begin{verbatim}
[px,py] = meshgrid([0:0.01:1]);

>>px(1:3,1:3)

ans =
         0    0.0100    0.0200
         0    0.0100    0.0200
         0    0.0100    0.0200
         ...
         
>>py(1:3,1:3)

ans =
         0         0         0
    0.0100    0.0100    0.0100
    0.0200    0.0200    0.0200
    ...

mu=cos(10*px)+sin(8*py);
e=normrnd(0,0.4,101,101); 
Y=mu+e;
figure; imagesc(Y); colorbar;
\end{verbatim}

The simulated image $Y$ is fairly noisy. Gaussian kernel smoothing with bandwidth 1 is applied to $Y$ to increase smoothness. The Gaussian kernel is constructed using {\tt inline} function. 
\begin{verbatim}
>>K=inline('exp(-(x.^2+y.^2)/2/sig^2)')

K =
     Inline function:
     K(sig,x,y) = exp(-(x.^2+y.^2)/2/sig^2)
\end{verbatim}
The inline function {\tt K} has three arguments: two coordinates {\tt x} and {\tt y} and bandwidth {\tt sigma}. The kernel is then constructed discretely using $5 \times 5$ grid and then renormalizing it such that the kernel sums up to 1. 
 For small {\tt sig}, the discrete kernel weights are focused in the center of the $5 \times 5$  window while for large {\tt sig}, the kernel weights are more dispersed. Error will increase if a smaller window is used. Smoothing is done by the convolution {\tt conv2}. The smoothed image is in Figure \ref{fig:kernelsmoothing}.

\begin{verbatim}
[dx,dy]=meshgrid([-2:2]);
>>weight=K(0.5,dx,dy)/sum(sum(K(0.5,dx,dy)))

weight =
0.0000 0.0000 0.0002 0.0000 0.0000
0.0000 0.0113 0.0837 0.0113 0.0000
0.0002 0.0837 0.6187 0.0837 0.0002
0.0000 0.0113 0.0837 0.0113 0.0000
0.0000 0.0000 0.0002 0.0000 0.0000

>>weight=K(1,dx,dy)/sum(sum(K(1,dx,dy)))

weight =
0.0030 0.0133 0.0219 0.0133 0.0030
0.0133 0.0596 0.0983 0.0596 0.0133
0.0219 0.0983 0.1621 0.0983 0.0219
0.0133 0.0596 0.0983 0.0596 0.0133
0.0030 0.0133 0.0219 0.0133 0.0030

Ysmooth=conv2(Y,weight,'same');
figure; imagesc(Ysmooth); colorbar;
\end{verbatim}

\section{Case Study: Smoothing of DWI Stroke Lesions}
One slice of sample T1 MRI is stored as bitmap image format as {\tt DWI.bmp}, which can be read using the built-in MATLAB function {\tt imread}. Stroke lesions in DWI are segmented by a human expert and stored as\\
{\tt DWI-segmentation.bmp}.

\begin{figure}[t]
\includegraphics[width=0.9\linewidth]{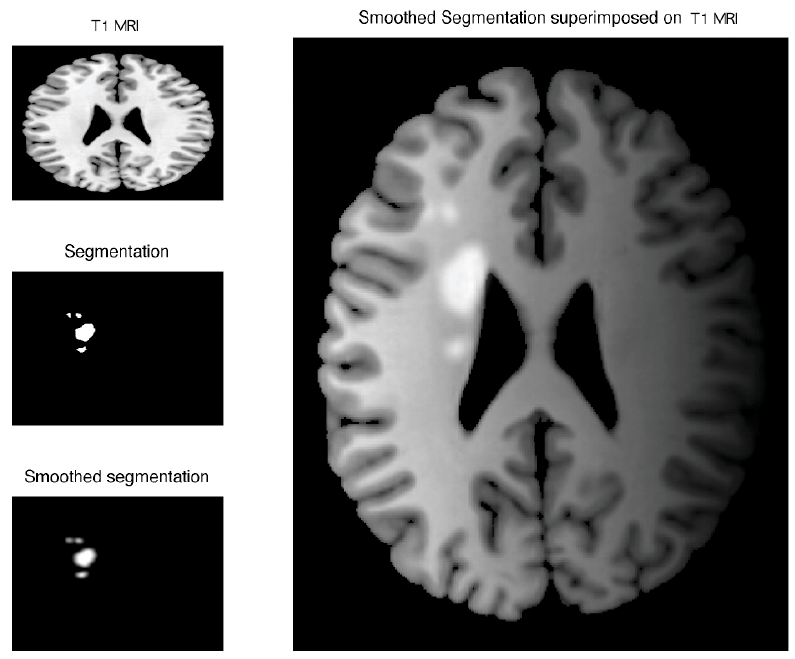}
\centering
\caption{T1 MRI and the segmentation of ischemic regions from diffusion-weighted image (DWI). The  segmentation is done manually using DWI and smoothed with Gaussian kernel. Smoothing is necessary before any group analysis in reducing the regions of false positives. The images are obtained as a part of study in \citet{ryu.2014}.}
\label{fig:smoothing-DWI}
\end{figure}

\begin{verbatim}
DWimg= 'DWI';
DWI=imread(DWimg,'bmp');

segmentation = 'DWI-segmentation';
f=imread(segmentation,'bmp');
\end{verbatim}

Once we read the image files, we are interested in smoothing the binary segmentation using the Gaussian kernel smoothing procedure. Bitmap files consist of a matrix of three columns representing red, green and blue colors. So we simply take out the first column and normalize it to have pixel values 0 or 1. The normalized segmentation is then smoothed using Gaussian kernel smoothing, which is implemented as {tt gaussblur2D.m}. The first argument to function {\tt gaussblur2D} is the 2D array to smooth out and the second argument is the bandwidth given in terms of full width at the half maximum (FWHM) of the kernel. 

\begin{verbatim}
binarize = double(f(:,:,1)/255);
fblur= gaussblur2D(binarize,10);  

n=size(fblur);  
fblurout = zeros(n(1), n(2), 3); 
fblurout(:,:,1)=fblur;
fblurout(:,:,3)=fblur;
\end{verbatim}

For visualization, we use {\tt subplot} consisting of $3 \times 3$ image grid. 
Four different images are displayed by partitioning the $3 \times 3$ image grid. 
We have superimposed the smoothed segmentation on top of MRI by specifying the transparency of each pixel using {\tt set} command with  {\tt 'AlphaData'} option. Figure \ref{fig:smoothing-DWI} is obtained by running the codes below.

\begin{verbatim}
figure; subplot(3,3,1);
imagesc(DWI);
axis off
title('DWI')

subplot(3,3,4);
imagesc(f);
axis off
title('Segmentation')

subplot(3,3,7); imagesc(fblurout)
axis off
title('Smoothed segmentation')

subplot(3,3,[2:3, 8:9]);
imagesc(DWI);
axis off
hold on

imgAlpha = repmat(0:1/n(2):1-1/n(2),n(1),1);
img = imagesc(fblurout);
set(img,'AlphaData',imgAlpha);
title('Smoothed Segmentation superimposed on DWI')
\end{verbatim}

\section{Checking Gaussianness}

In this section, we will explain various methods for checking Gaussianness of imaging measurements. Since many statistical models assume normality, checking if imaging data follows normality is fairly important. Also image smoothing not only increases the smoothness of underlying image intensity values, but it also increases the normality of data. In paper submissions to various imaging journals, this is an often asked question by reviewers.  

\subsection{Quantitle-Quantile Plots}
\index{quantile-quantile plots}
\index{quantile}

Checking the normality of imaging data is fairly important when the underlying statistical model assumes the normality of data. But how do
we know the data will follow normality? This is easily checked visually using the {\em quantile-quantile (QQ) plot} first introduced by Wilk and Gnanadesikan \citep{wilk.1968}. The QQ-plot is a graphical method for comparing two distributions by plotting their quantiles against each other. A special case of QQ-plot is the {\em normal probability plot} where the quantiles from an empirical distribution are plotted on the vertical axis while the theoretical quantiles from a Gaussian distribution are plotted on the horizontal axis. It is used to check graphically if the empirical distribution follows
the theoretical Gaussian distribution. If the data follows Gaussian, the normal probability plot should be close to a straight line.\\

\subsection{Quantiles} 
\index{quantile}

\begin{definition}
The quantile point $q$ for random variable $X$ is a point that satisfies 
$$P(X \leq q) =F_X(q)= p,$$ where $F_X$ is the cumulative distribution function
(CDF) of $X$. 
\end{definition}

Assuming we can find the inverse of CDF, the quantile is given by
$$q=F_X^{-1}(p).$$ 
This function is mainly referred to as a quantile function. 
The quantile-quantile (QQ) plot of two random variables $X$ and $Y$ is then defined to be a parametric curve $\mathcal{C}(p)$ parameterized by $p \in [0,1]$:
$$\mathcal{C}(p) = \big(F_X^{-1}(p), F_Y^{-1}(p)\big).$$

\subsection{Empirical Distribution} 
\index{distributions ! empirical}
\index{distributions ! cumulative}
\index{empirical distribution}
\index{cumulative distribution}

The CDF $F_X(q)$ measures the proportion of random variable $X$ less than given value $q$. So by counting the number of measurements less than $q$, we can empirically estimate the CDF. Let $X_1,\cdots, X_n$ be a random sample of size $n$. Then order them in increasing order:
$$ \min (X_1,\cdots,X_n) = X_{(1)} \leq X_{(2)} \leq \cdots \leq X_{(n)}  = \max (X_1,\cdots, X_n).$$
Suppose $X_{(j)} \leq q < X_{(j+1)}$. This implies that there are  $j$ samples that are smaller than $q$. 
So we approximate the CDF as
$$\widehat{F_X}(q) = \frac{j}{n}.$$
The $j/n$-th {\em sample quantile} is then $X_{(j)}$. Some authors define the sample quantile as the $(j-0.5)/n$-th sample quantile. The factor 0.5 is introduced to account for the descritization error.

In numerical implementation, it is easier to implement the empirical distribution using the {\em step function} $\mathcal{I}_q(x)$ which is implemented as
$\mathcal{I}_q(x) = 1$ if $ x\leq q$ and $\mathcal{I}_q(x) = 0$ if $ x> q$. Then the CDF is estimated as
$$\widehat{F_X}(q) = \frac{1}{n} \sum_{i=1}^n \mathcal{I}_q(X_i),$$
where $\mathcal{I}_q(X_i)$ counts if $X_i$ is less than $q$.  A different possibly more sophisticated estimation can be found in \citep{frigge.1989}.

\subsection{Quantile-Quantile Plots} 
\index{quantile-quantile plots}

\begin{figure}[t]
\begin{center}
\includegraphics[width=0.9\linewidth]{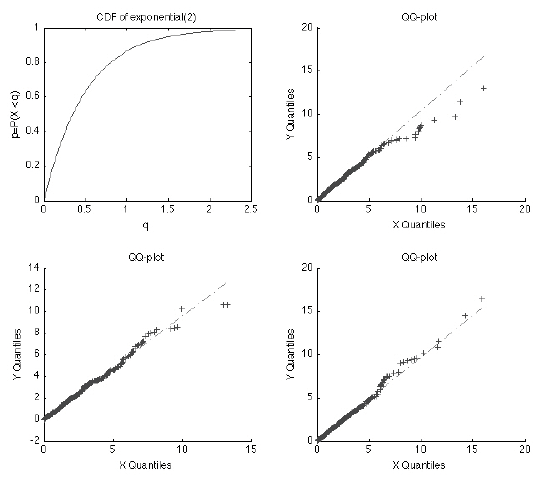}
\caption{Top left: CDF of a exponential random variable with parameter 2. QQ-plots are obtained by randomly generating 500 exponential random variables. Each time, we may have slightly different QQ-plots at extreme values. In interpreting QQ-plots, we should not focus attention on extreme outlying points.}
\label{fig:quantitle-exponential}
\end{center}
\end{figure}

The QQ-plot for two Gaussian distributions is a straight line. This can be easily proved as follows. Suppose $$X \sim N(\mu_1,\sigma_1^2) \mbox{ and } Y \sim N(\mu_2,\sigma_2^2).$$ Let $Z \sim N(0,1)$ and $\Phi(z) =P(Z \leq z),$ the CDF of the standard normal distribution. If we denote $q_1$ and $q_2$ to be the $p$-th quantiles for $X$ and $Y$ respectively, we have
$$p = P(X \leq q_1) = P\Big(\frac{X-\mu_1}{\sigma_1} \leq \frac{q_1-\mu_1}{\sigma_1}  \Big)=\Phi \Big(\frac{q-\mu_1}{\sigma_1}  \Big).$$ 
Hence the parameterized QQ-plot is given by 
\bq q_1(p) &=& \mu_1 + \sigma_1\Phi^{-1}(p),\\
q_2(p) &=& \mu_2 + \sigma_2\Phi^{-1}(p).\eq
This is a parametric form of the QQ-plot. The QQ-plot without the parameter $p$ is  then trivially given by
$$\frac{q_1 -\mu_1}{\sigma_1} = \frac{q_2 -\mu_2}{\sigma_2},$$
the equation for a line. This shows the QQ-plot of two normal distributions is a straight line. This idea can be used to determine the normality of a given sample. We can check how closely the sample quantiles correspond to the normal distribution by plotting the QQ-plot of the sample quantiles vs. the corresponding quantiles of a normal distribution. In normal probability plot, we plot the QQ-plot of the sample against the standard normal distribution $N(0,1)$. 

\subsection{{\tt MATLAB} Implementation}
As an example, consider the problem of plotting the quantile function for the exponential random variable $X$ with parameter $\lambda = 2$, i.e. $X \sim exp(2)$. It can be shown that
$$F_X^{-1}(p) =
-\frac{1}{2}\ln(1-p).$$
The actual CDF can be plotted using the {\tt inline} function, which can define a function quickly without writing a separate function file. 
\begin{verbatim}
p=[1:99]/100; 
q=inline('-log(1-p)/2');
plot(q(p),p);
xlabel('x')
ylabel('P(X < q)
\end{verbatim}

When we generate the QQ plot of $X \sim exp(2)$ and $Y \sim
exp(2)$, since they are identical distributions, you expect the straight line
$y=x$ as the QQ plot (Figure \ref{fig:quantitle-exponential}). This can be done by the exponential random number generator {\tt exprnd}. 
\begin{verbatim}
X=exprnd(2, 500,1);
Y=exprnd(2, 500,1);
subplot(2,2,2); qqplot(X,Y)
title('QQ-plot')
\end{verbatim}

\section{Effect of Gaussianness on Kernel Smoothing}
\index{Gaussianness}

\begin{figure}[t]
\centering
\includegraphics[width=0.9\linewidth]{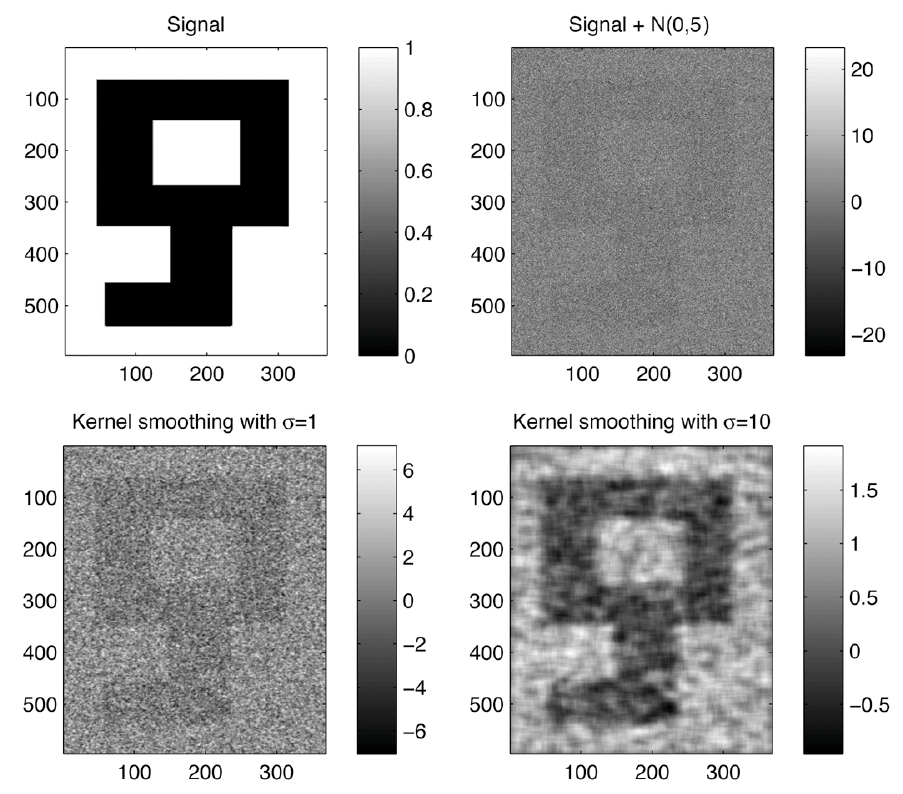}
\caption{If properly used, Gaussian kernel smoothing can be used in detecting a hidden signal. The key shaped signal is contaminated with $N(0,5)$. Since the noise variability is so huge, the underlying signal is not clearly visible. By performing kernel smoothing with increasing bandwidths, it is possible to recover the original signal.   
}\label{fig:kernel-1Dkey}
\end{figure}

Gaussian kernel smoothing can increase the Gaussianness of multiple images. As an example, consider a key shaped binary image {\tt toy-key.tif} (Figure \ref{fig:kernel-1Dkey}). Gaussian noise with large variance is added to the binary image to mask the signal. 

\begin{verbatim}
signal= imread('toy-key.tif');
signal=(double(signal)-219)/36; 
figure; subplot(2,2,1); imagesc(signal); colormap('bone'); colorbar 
title('Signal')

noise= normrnd(0, 5, 596, 368);
f = signal + noise;
subplot(2,2,2); imagesc(f); colormap('bone'); colorbar
title('Signal + N(0,5)')
\end{verbatim}

To recover the signal, we performed Gaussian kernel smoothing with the bandwidths 1 and 10. For a sufficiently large bandwidth of 10, we are able to recover the underlying key shaped object. If properly used, Gaussian kernel smoothing can recover the underlying signal pretty well.

\begin{figure}[t]
\centering
\includegraphics[width=0.9\linewidth]{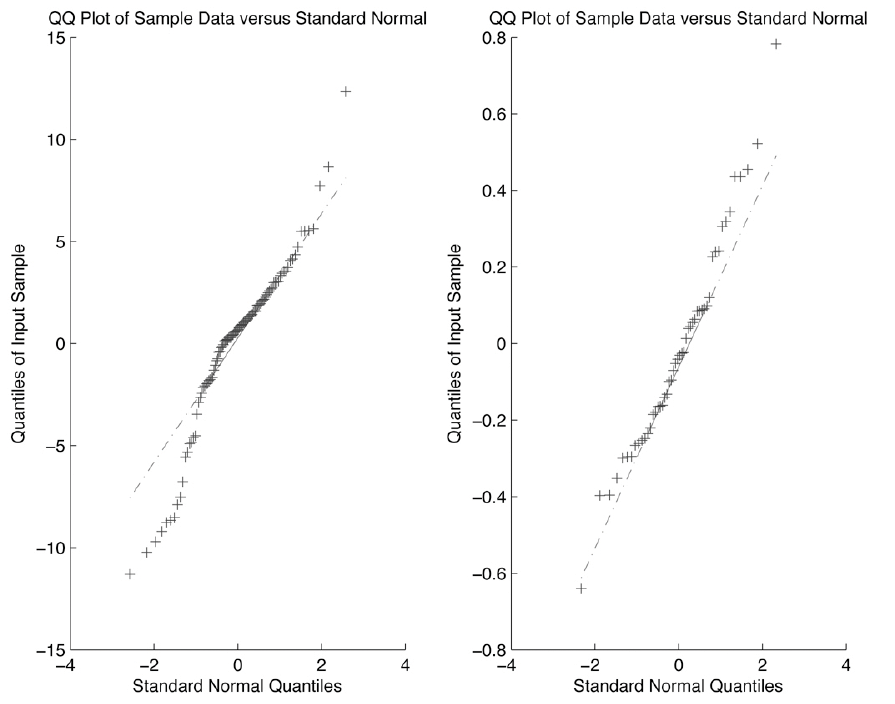}
\caption{Image intensity value at a particular pixel. After Gaussian kernel smoothing with bandwidth 100, Gaussianness has increased as expected.}\label{fig:kernel-QQplot}
\end{figure}

\begin{verbatim}
K=inline('exp(-(x.^2+y.^2)/2/sigma^2)'); 
[dx,dy]=meshgrid([-10:10]);  

sigma=100;
weight=K(sigma,dx,dy)/sum(sum(K(sigma,dx,dy)));

weight=K(1,dx,dy)/sum(sum(K(1,dx,dy)));
smooth=conv2(f,weight,'same');
subplot(2,2,3); imagesc(smooth); colormap('bone'); colorbar
title('Kernel smoothing with \sigma=1')

weight=K(10,dx,dy)/sum(sum(K(10,dx,dy)));
smooth=conv2(f,weight,'same');
subplot(2,2,4); imagesc(smooth); colormap('bone'); colorbar
title('Kernel smoothing with \sigma=10')
\end{verbatim}

To show kernel smoothing can increase Gaussianness, we selected a pixel at $(314, 150)$ which is at the edge of the binary object. For 50 measurements at the pixel, we plotted the QQ-plot. As shown in the left in Figure \ref{fig:kernel-QQplot}, the pixel values are not showing Gaussianness. However, after smoothing with bandwidth 100, the Gaussianness has been increased.

\begin{verbatim}
for i=1:50
    noise= normrnd(0, 5, 596, 368);
    weight=K(100,dx,dy)/sum(sum(K(100,dx,dy)));
    f=signal+noise;
    smooth=conv2(f,weight,'same');
    pixelvalue(i)=f(314, 150);
    pixelvalues(i)=smooth(314, 150);
end;

figure; subplot(1,2,1); qqplot(pixelvalue)
subplot(1,2,2); qqplot(pixelvalues)

\end{verbatim}

\section{Relation to kernel density estimation}
Given scatter points $x_1, \cdots, x_n \in \mathbb{R}^d$ as observation, their empirical distribution is given by  
$$f(x) = \frac{1}{n}\sum_{i=1}^n \delta(x - x_i).$$
We have impulse functions at scatter points such that 
$$\int_{\mathbb{R}^d}  f(x) \; dx = 1.$$ 
Thus, $f$ is a probability density that is somewhere between continuous and discrete. The Dirac delta function is often used in converting discrete data into continuous functional form, which provide easier algebraic manipulation. If we apply Gaussian kernel smoothing with kernel $K_{\sigma}$, we have
\bqn K_{\sigma} * f(x)  = \frac{1}{n}\sum_{i=1}^n K_{\sigma} (x-x_i),\eqn
which is exactly the kernel density estimation often encountered in statistical literature \citep{fan.1996}.

\section*{Acknowledgements}

We would like to thank  Dong-Eog Kim of  Dongguk University Ilsan Hospital for providing DWT data used as a case study.
We would like to thank students in various image analysis classes at both University of Wisconsin-Madison and Seoul National University who used the materials in this paper. This study is funded by NIH R01 EB02875 and NSF MDS-2010778.

\bibliographystyle{agsm} 
\bibliography{reference.2020.07.15}

\end{document}